\documentclass[aps,pra,epsfigure,twocolumn,showpacs]{revtex4}%{revtex4-1}
\usepackage{amsmath}
\usepackage{amstext}
\usepackage{latexsym}
\usepackage{graphicx}
\usepackage{amsfonts}
\usepackage{epsfig}
\usepackage{color}

\begin{document}
\title{Deterministic implementation of weak quantum cubic nonlinearity}
\author{Petr Marek}
\affiliation{Department of Optics, Palack\'{y} University, 17. listopadu 1192/12,
77146 Olomouc, Czech Republic}
\author{Radim Filip}
\affiliation{Department of Optics, Palack\'{y} University, 17. listopadu 1192/12,
77146 Olomouc, Czech Republic}
\author{Akira Furusawa}
\affiliation{Department of Applied Physics, School of Engineering, The University of Tokyo, 7-3-1 Hongo, Bunkyo-ku, Tokyo 113-8656, Japan}

\begin{abstract}
We propose a deterministic implementation of weak cubic nonlinearity, which is a basic building block of a full scale CV quantum computation. Our proposal relies on preparation of a specific ancillary state and transferring its nonlinear properties onto the desired target by means of deterministic Gaussian operations and feed-forward. We show that, despite the imperfections arising from the deterministic nature of the operation, the weak quantum nonlinearity can be implemented and verified with the current level of technology.
\end{abstract}

\maketitle

Ever since it has been first mentioned by Feynman \cite{Feynman}, quantum computation has been the holy grail of quantum information theory, because the exponential speedup it offers promises a significantly better way to tackle certain non-polynomial computational tasks. However, in order to achieve this, high-order non-linear resources are needed. The original approach to quantum computing relied on manipulation of quantum systems with a limited Hilbert space \cite{KLM}, but one can expand this concept to arbitrary quantum systems and say that the universal quantum computation is defined by its ability to emulate a Hamiltonian in form of an arbitrary (Hermitian) polynomial of annihilation and creation operators \cite{Lloyd99}.

%At first glance this may seem as an insurmountable obstacle, but it is a celebrated result of Lloyd and Braunstein \cite{Lloyd99} that if one has access to two kinds of operations governed by two different Hamiltonians, one can use these operations in sequence to implement a different operation governed by the commutator of the two Hamiltonians.

It is a great advantage of the continuous variables (CV) quantum systems \cite{CVQIP} that Hamiltonians composed of first (linear) and second (quadratic) powers of quadratic operators $\hat{x}$ and $\hat{p}$ can be readily implemented. However, it can be easily checked that these are not enough and that an access to cubic Hamiltonian with at least third power of quadrature operators is necessary. Unfortunately, the currently achievable experimental interaction strengths are too low compared to noise to be of use.

Fortunately, the need for currently unavailable cubic unitary evolution may not be so dire. Let us recall the original statement of Lloyd and Braunstein \cite{Lloyd99}: If one has access to Hamiltonians $\hat{A}$ and $\hat{B}$, one can approximatively implement operation with Hamiltonian $i[\hat{A},\hat{B}]$. \emph{Approximatively} is the key term here, meaning that the desired operation is engineered only as a quadratic polynomial of the interaction time:
\begin{equation}\label{commutrick}
e^{i\hat{A}t} e^{i\hat{B}t} e^{-i\hat{A}t} e^{-i \hat{B}t} \approx e^{-[\hat{A},\hat{B}]t^2} + O(t^3).
\end{equation}
Consequently, even the initial operations need not to be unitary - their quadratic approximations are fully sufficient. What this means is that if we take interest in a sample cubic interaction with Hamiltonian $H\propto x^3$, we need not to implement the unitary $e^{i\chi x^3}$, but it is enough to be able to perform operation
\begin{equation}\label{approximate}
   \mathcal{O}_{6}(\hat{x}) =  1+i\chi \hat{x}^3 - \chi^2 \hat{x}^6/2.
\end{equation}
This is the lowest order expansion for which the commutator trick (\ref{commutrick}) works, but let us start with the real lowest order expansion, $1 + \beta \hat{x}^3$, where $\beta$ is a complex number. This expansion behaves as weak cubic coupling if beta is imaginary and has the added benefit that is can be used to compose (\ref{approximate}) when the respective values of $\beta$ are complex and chosen properly. In principle, even this gate can be further decomposed into series of $1+ \gamma \hat{x}$ ($\gamma \in \mathbb{C}$) operations \cite{filipmarekprep}. These phase sensitive gates can be implemented probabilistically on a traveling beam of light by subsequent application of photon subtraction and photon addition, represented by operators $\hat{a} = (\hat{x} + i\hat{p})/\sqrt{2}$ and $\hat{a}^{\dag}$ \cite{Bellini,Fiurasek09,Bellini2}. They are very useful for preparing various ancillary states, but for use in a full-fledged information processing we are interested in their \emph{deterministic} implementation.

\textbf{Ideal implementation:} To this end we employ the approach of \cite{Gottesman}, thoroughly discussed in \cite{Ghose}, where it was suggested that unitary operation acting on a state can be deterministically implemented with help of a proper resource state, a QND coupling, a suitable measurement and a feed-forward loop. Explicitly, for operation $\mathcal{O}(\hat{x})$ acting on pure state $|\psi\rangle = \int \psi(x)|x\rangle dx$ the resource state is $\mathcal{O}(\hat{x})|p=0\rangle$. After QND coupling, represented by unitary $\hat{U}_{\mathrm{QND}}(\lambda) = e^{i \lambda \hat{x}_2 \hat{p}_1}$, is employed and the overall state is transformed to
\begin{equation}\label{}
    \int \psi(x) \mathcal{O}(y) |y-\lambda x,x\rangle dx dy,
\end{equation}
the ancillary resource mode gets measured by a homodyne detection. We can for now assume $\lambda = 1$, as the overall message remains unchanged. For any detected value $q$ the output state is
\begin{equation}\label{gate}
    \int\psi(x) \mathcal{O}(x+q)|x\rangle dx.
\end{equation}
To obtain the desired result, one either post-select only for situations when $q=0$ was detected, or apply a feed-forward which would compensate for $x+q$ in the argument of the operator. It has been shown in \cite{Gottesman} that if the desired operation $\mathcal{O}(x)$ is a unitary operation driven by a Hamiltonian of order $n$, the feed-forward operation requires Hamiltonian of order $n-1$. Explicitly, imperfections in operator $\mathcal{O}(\hat{x}+q) = \exp[i\chi(\hat{x}+q)^3]$ can be compensated by unitary operator $\hat{U}_{\mathrm{FF}} = \exp[-i\chi (3q\hat{x}^2+ 3q^2\hat{x})]$, which is a combination of displacement, squeezing and phase-shifts. The operation (\ref{approximate}) we are interested in is not unitary, but since it is an approximation of unitary driven by a cubic Hamiltonian, a feed-forward of squeezing and displacements should perform adequately, up to some error. We'll get to this issue later. In fact, the operations available for feed-forward limit us in what we can do. With squeezing and displacement we can implement only cubic operations. Of course, with them we could also tackle Hamiltonians of the fourth order, and so on. And there is another limitation - since the feed-forward must be deterministic and noiseless, and therefore unitary, it can be only used to deterministically compensate unitary (at least approximatively) operations whose Hamiltonian is hermitian. Therefore we cannot use the trick with implementing a series of $1+\gamma \hat{x}$ operations, we have to implement operation (\ref{approximate}) in one go. Consequently, we need a sufficiently complex resource state.

\textbf{Resource state generation:} Let us now shift our attention to the required resource state. In realistic, even if idealized, considerations one has to, instead of a position eigenstate, use a squeezed state $S|0\rangle = [\int \exp(-x^2/g)|x\rangle dx]/(\pi g)^{1/4}$, which approaches the ideal form as $g\rightarrow\infty$. The resource state can now be expressed as:
$ \mathcal{O}(\hat{x})\hat{S}|0\rangle = \hat{S}\mathcal{O}(\hat{x}/\sqrt{g})|0\rangle $
which is a state finite in Fock basis with superficial layer of squeezing. As it has a finite structure, the state can be engineered by a sequence of six photon additions \cite{Dakna} or photon subtractions \cite{Fiurasek05}. This is an extremely challenging task, let us therefore first focus at the lowest nontrivial cubic hamiltonian expansion, $\mathcal{O}_{3}(\hat{x}) = 1+ \chi x^3$, which is a feasible extension of recent experimental works \cite{subtractionexp}. The appropriate resource state looks as
\begin{equation}\label{chistate}
\hat{S} (1+ \chi' \hat{x}^3)|0\rangle =
\hat{S}\left( |0\rangle + \chi'\frac{3}{2\sqrt{2}}|1\rangle + \chi'\frac{\sqrt{3}}{2}|3\rangle\right),
\end{equation}
with $\chi' = \chi g^{-3/2}$. This state can be generated from a squeezed state by a proper sequence of photon subtractions and displacements \cite{Fiurasek05}, which acts as $(\hat{a} -\alpha)(\hat{a}-\beta)(\hat{a}-\gamma)\hat{S}|0\rangle$.
Since the squeezing operation transforms the annihilation operator as $\hat{S}^{\dag}\hat{a}\hat{S} = \mu \hat{a} - \nu \hat{a}^{\dag}$, where $\mu = \cosh(\ln\sqrt{g})$ and $\nu = \sinh(\ln\sqrt{g})$, the required displacements can be obtained as a solution of set of equations:
\begin{eqnarray}\label{equations}
  A = \alpha\beta\gamma, ~\alpha + \beta +\gamma = 0, \nonumber \\
  2\sqrt{2} \nu^3 =  A \chi' , ~3\nu^2+ 3\mu\nu = \left(\alpha\beta + \alpha\gamma + \beta\gamma\right),
\end{eqnarray}
where $A$ is a constant parameter related to normalization. The solution exists and it can be found analytically as
\begin{eqnarray}
  \alpha &=& \frac{\xi + \sqrt{\xi^2-4\zeta}}{2}, \nonumber \\
  \beta &=& \frac{\xi - \sqrt{\xi^2-4\zeta}}{2}, \nonumber \\
  \gamma &=& -(\alpha + \beta).
\end{eqnarray}
Here $\xi$ and $\zeta$ are solutions of the set of equations
\begin{eqnarray}\label{cubiceq}
  xy+C_1 = 0 , \quad  y-x^2 - C_2 = 0,
\end{eqnarray}
where $C_1 = \nu^3 2\sqrt{2}\chi'^{-1}, \quad C_2 = 3\nu^2 + 3\mu\nu$.
The solutions of (\ref{cubiceq}) always exist and they can be obtained analytically using the Cardan formula.
%\begin{eqnarray}\label{}
% x_{1,2,3} = \hspace{7cm}\\ \left(-\frac{C_1}{2}+\sqrt{\frac{C_1^2}{4}+\frac{C_2^3}{27}}\right)^{1/3} + \left(-\frac{C_1}{2}-\sqrt{\frac{C_1^2}{4}+\frac{C_2^3}{27}}\right)^{1/3}, \nonumber
%\end{eqnarray}
%here the particular cube roots are chosen so that their product is equal to $C_2/3$.
The squeezing used in the state generation can be in general different from the squeezing in (\ref{chistate}). However, squeezing can be considered to be a well accessible operation and we shall therefore not deal with this in detail. It should be noted that an alternative way of preparing the state (\ref{chistate}) lies in performing a suitable projection onto a single mode of a two-mode squeezed vacuum state. Engineering of the proper measurement, which too requires three APDs and three displacements, leads to similar equations as in the previous case (\ref{equations}) with solution of the same form.

\textbf{Realistic implementation:} With the resource state at our disposal we can now look more closely at the two ways to implement the gate, the probabilistic and the  deterministic, in order to compare them and see, what is the manifestation of high order nonlinearity in the deterministic case. The probabilistic implementation is rather straightforward. Using the resource state (\ref{chistate}) we are able to transform the initial state to
\begin{equation}\label{psi_prob}
    |\psi_{0}\rangle = \int \psi(x) \mathcal{O}(x) e^{-x^2/2g}|x\rangle dx,
\end{equation}
and as the squeezing of the resource state approaches infinity, the produced state approaches its ideal form. The final state is always pure and the actual composition of the operator $\mathcal{O}_n(x)$ can be arbitrary, allowing us, for example, to implement operator $\mathcal{O}$ in $n$ different non-unitary steps. On the other hand, if the resource squeezing is insufficient compared to the distribution of the state in phase space, it seriously affects some properties of the state - for example, moments of $x$ quadrature may not be preserved any more.

But let us move towards the more interesting part, the deterministic approach. In this case the operation produces a mixed state
\begin{equation}\label{}
    \rho' = \int P(q) |\psi_q\rangle\langle\psi_q|dq.
\end{equation}
Here, $P(q)$ represents the probability of measuring a specific outcome $q$ and
\begin{eqnarray}\label{}
    |\psi_q\rangle = \frac{1}{\sqrt{P(q)}\mathcal{N}_R} \times \hspace{4cm} \\
    \int \psi(x) e^{-(x+q)^2/2g} \mathcal{O}(x+q) e^{-i\chi q^3 -i3\chi (x q^2 + x^2 q)}  |x\rangle dx, \nonumber
\end{eqnarray}
where $\mathcal{N}_R$ is the norm of the resource state, stands for the respective quantum state corrected by feed-forward. Ideally, $\mathcal{O}(x+q) e^{-i3\chi (x q^2 + x^2 q)} \approx \mathcal{O}(x)$, but this relation can obviously work only when both $x$ and $q$ are small enough for the exponent to be reasonably approximated by the finite expansion $\mathcal{O}_n$. It is therefore quite unfortunate that the very condition required for the operation to work flawlessly, the need for $g\rightarrow \infty$, is compatible with the feed-forward only in the limit of $\chi\rightarrow 0$. To quantify these properties in greater detail we need to employ a suitable figure of merit.

\textbf{Analysis:} To evaluate the quality of the approximate operation is not a straightforward task. 
%A traditional approach is the quantum process tomography \cite{gatetomography}, which applies the operation under question to one part of the maximally entangled state and compares the real situation with the ideal one by means of fidelity. However, this method is not directly applicable to our scenario. There are two main reasons for that: one, the maximally entangled CV state possesses infinite energy while the approximate method is expected to work only for weak signals; two - the figure of merit, fidelity, is mostly concerned with gross features of the quantum states and has problems in differentiating between weakly non-Gaussian states, that are generated by the approximate gate, and Gaussian states with just enough similarity.
If we want to conclusively distinguish the cubic type nonlinear interaction from a Gaussian one, we can take advantage of the known way the quadrature operators transform: $\hat{x} \rightarrow \hat{x}$, $\hat{p}\rightarrow \hat{p}+\chi \hat{x}^2$. If we apply the operation, in form of a black box, to a set of known states, we can analyze the transformed states to see whether the operation could be implemented by a suitable Gaussian, or if it is more of what we aim for. The analysis can be as easy as checking the first two moments of the quadrature operators, because the nonlinear dependance of $\langle \hat{p}\rangle$ on $\langle \hat{x}\rangle$ can not be obtained by a Gaussian operation, unless we consider a rather elaborate detection-and-feed-forward setup, which would, however, introduce an extra noise detectable either by checking purity of the state, or by analyzing higher moments $\langle \hat{x}^2\rangle$ and $\langle \hat{p}^2\rangle$.

The case with purity of one is straightforward to verify - as soon as the first moments have the desired form, $\langle \hat{x}'\rangle = \langle \hat{x}\rangle$ and $\langle \hat{p}'\rangle = \langle \hat{p}\rangle + \chi \langle \hat{x}^2\rangle$, we can be certain a form of the desired non-Gaussianity is at play. In the presence of noise, the confirmation process is more involved and we shall deal with it in a greater detail.

It needs to be shown that, in comparison to the deterministic approximation, no Gaussian operation can provide the same values of moments $\langle \hat{x}'\rangle$, $\langle \hat{x}'^2\rangle$, and $\langle \hat{p}'\rangle$ without also resulting in a significantly larger value of moment $\langle \hat{p}'^2\rangle$ caused by the extra noise. The complete Gaussian scheme consists of an arbitrary Gaussian interaction of the target system with a set of ancillary modes followed by a set of measurements of these modes yielding values which are used in suitable feed-forward to finalize the operation. In the case where the approximate transformations approach the ideal scenario, i.e. when $\langle \hat{x}'\rangle = \langle \hat{x}\rangle $, $\langle \hat{x}'^2\rangle = \langle \hat{x}^2\rangle$, and $\langle \hat{p}'\rangle = \langle \hat{p}\rangle + \chi \langle \hat{x}^2\rangle$, only a single ancillary mode is sufficient, the optimal Gaussian interaction is in the QND interaction with a parameter $\lambda$, and after a value of $\xi$ is measured by homodyne detection, feed-forward displacement of $\kappa \xi^2$ ensures the correct form of the three moments. In the end, the Gaussian approximated state can be expressed as
\begin{eqnarray}\label{benchmark}
 \rho''_{S} = \int dx \times \hspace{3cm} \\
 \hat{D}_S(\kappa x^2) _A\langle x|\hat{U}_{\mathrm{QND}}(\lambda) \hat{\rho}_S\otimes|0\rangle\langle 0| \hat{U}^{\dag}_{\mathrm{QND}}(\lambda) |x\rangle \hat{D}^{\dag}_S(\kappa x^2), \nonumber
\end{eqnarray}
where the subscripts $S$ and $A$ denote the signal and the ancillary mode, respectively. The high order classical nonlinearity is induced by the nonlinear feed-forward, represented by the displacement $\hat{D}(\alpha)$. The strength of the QND interaction $\lambda$ remains a free parameter over which can the procedure be optimized to obtain the best approximation characterized by the minimal possible value of the extra noise term in $\langle \hat{p}'^2\rangle $.

We analyze the aforementioned properties over a set of small coherent states $\alpha$ with $|\alpha|<2$. We compare the state obtained by the approximative cubic interaction with the state created by the Gaussian method. In principle, this could be done for both the deterministic and the probabilistic approach, but since the probabilistic approach has the potential to work perfectly, we shall keep to deterministic methods in our comparative endeavors. For each coherent state and its cubic-gate transformed counterpart we can, from knowledge of the first moments of quadrature operators, estimate the actual cubic nonlinear parameter and use it to construct the benchmark Gaussian-like state (\ref{benchmark}). The final step is to compare the extra noise present in $\hat{p}$ quadrature - if the added noise for the approximate state is below the Gaussian benchmark we can assume a non-Gaussian nature of the operation.
%%%%%%%%%%%%%%%%%%%%%%%%%%%%%%
\begin{figure}
\centerline{\psfig{figure=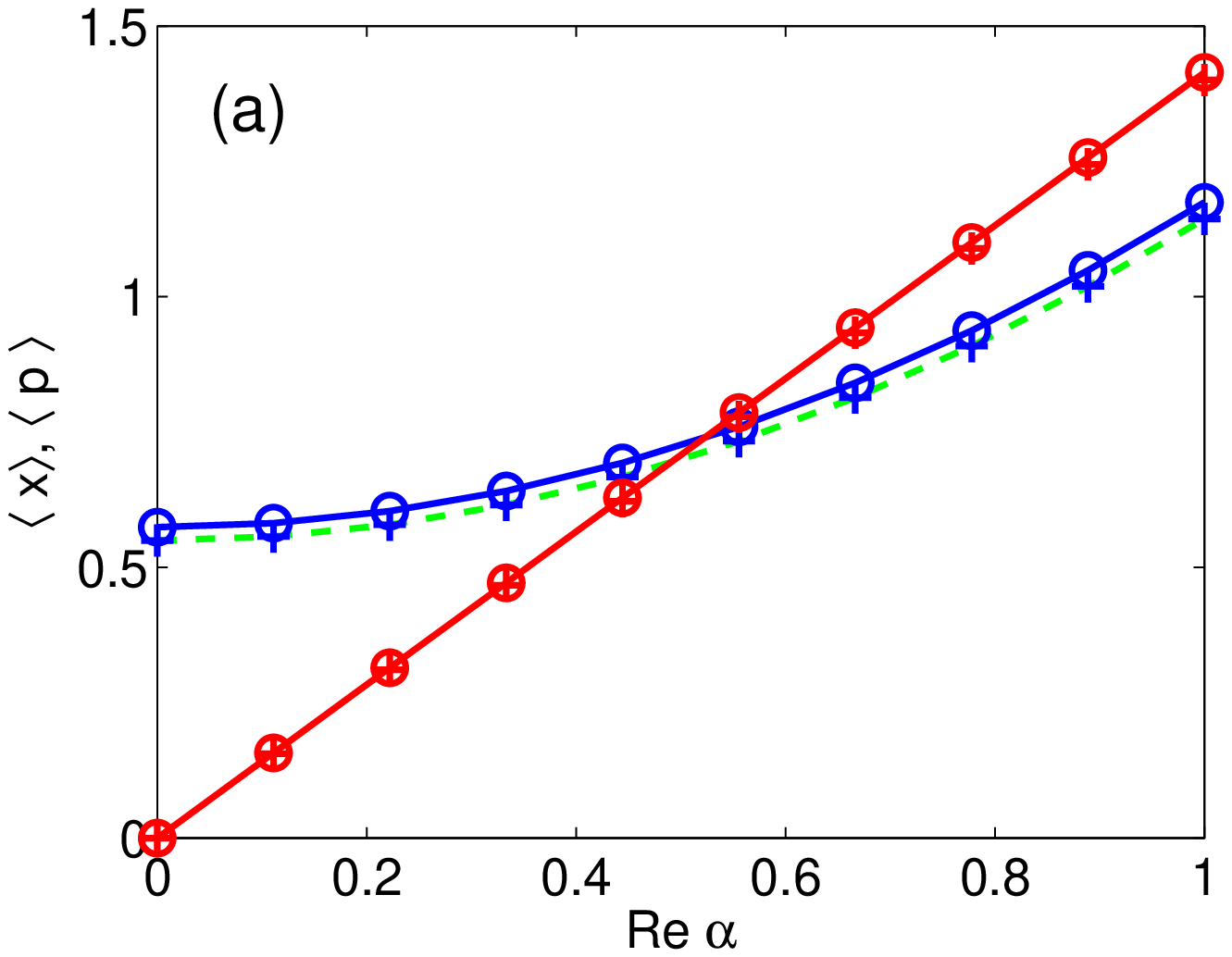,width=0.6\linewidth}}
\centerline{\psfig{figure=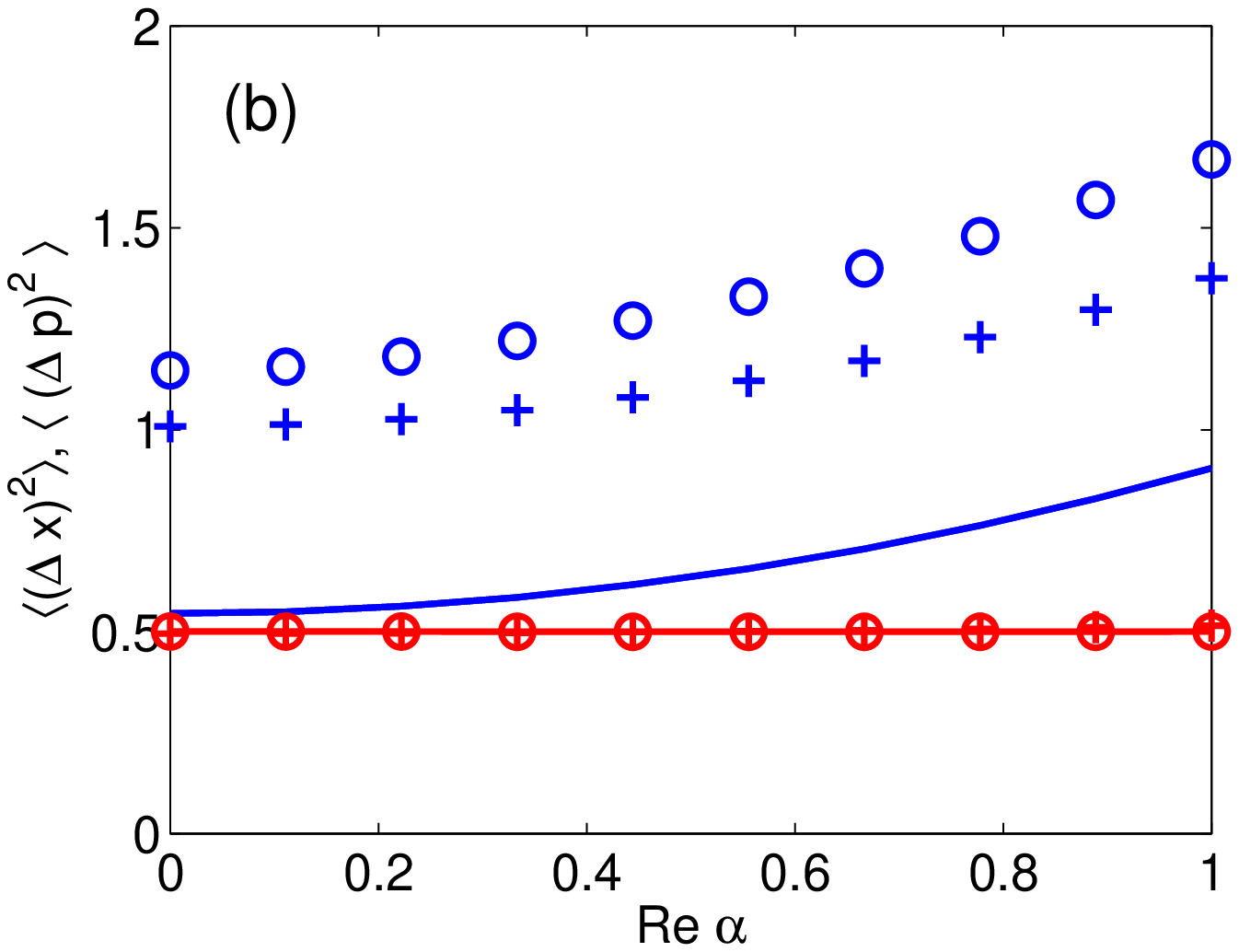,width=0.6\linewidth}}
\caption{(Color online) \textbf{(a)} First moments relative to real part of $\alpha$. Solid red and blue lines represent the ideal values of $\langle \hat{p}\rangle$ and $\langle \hat{x}\rangle$, respectively. Red and blue crosses than show these values for deterministic non-Gaussian approximation, while red and blue circles do so for the Gaussian approximation. Dashed green line is a quadratic fit for $\langle \hat{p}\rangle$. The experimental parameters are $g = 1$ and $\chi = 0.03$.
\textbf{(b)} Second moments relative to real part of $\alpha$. Solid red and blue lines represent the ideal values of $\langle \hat{p}^2\rangle$ and $\langle \hat{x}^2\rangle$, respectively. Red and blue crosses than show these values for deterministic non-Gaussian approximation, while red and blue circles do so for the Gaussian approximation.
} \label{fig_det}
\end{figure}
%%%%%%%%%%%%%%%%%%%%%%%%%%%%%%%%%%%%%%%%%%%%%%%%%%%

As an example, let us look at a particular scenario, in which the deterministic cubic gate was applied to a set of coherent states with imaginary part of the complex amplitude constant. The effect of the operation is illustrated in Fig.~\ref{fig_det}. Fig.~\ref{fig_det}a shows the first moments and reveals that for this purpose, effective cubic nonlinearity of $\chi_{\mathrm{eff}} = 0.1$ can be reliably obtained for both the non-Gaussian and the Gaussian approach. Differences arise, though, for the second moments, where the value of Gaussian quadrature moment $\langle \hat{p}^2\rangle$ is observably higher than the value of its non-Gaussian counterpart. The values of the Gaussian moment were obtained by optimizations of (\ref{benchmark}) for each particular value of $\mathrm{Re}\alpha$, it is therefore a stronger benchmark than a universal Gaussian operation, working over the whole range of $\mathrm{Re} \alpha$, would be. And it is still beaten by the imperfect deterministic non-Gaussian method with no squeezing in the ancillary mode.

\begin{figure}
\centerline{\psfig{figure=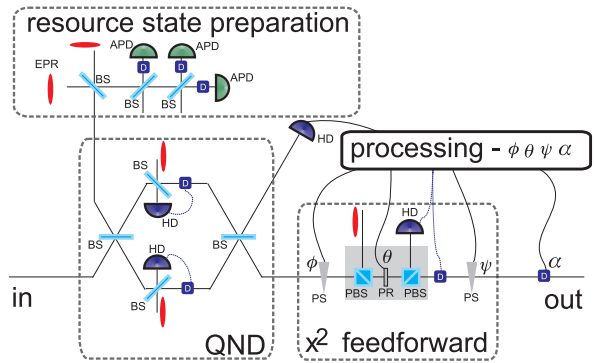,width=0.9\linewidth}}
\caption{(Color online) Schematic experimental setup of the deterministic $x^3$ gate. BS-beam splitters, PBS-polarization beam splitters, PS-phase shift, PR-polarization rotator, APD-avalanche photo diode, HD-homodyne detection, D-displacement. } \label{fig_setup}
\end{figure}

\textbf{Experimental setup} (Fig.~\ref{fig_setup})
%%%%%%%%%%%%%%%%%%%%%%%%%%%%%%%
%%%%%%%%%%%%%%%%%%%%%%%%%%%%%%%%%%%%%%%%%%%%%%%%%%%
The resource state generation requires three-photon subtraction from a squeezed, or a two-mode squeezed light, with appropriate displacements. Two photon subtractions have been already implemented \cite{subtractionexp} and three of them are within reach. The resource state is then coupled with the input using a QND gate with offline squeezing \cite{MIO_theory, QND_experiment}, which can be modified as to reliably manipulate the non-Gaussian resource state \cite{cat_teleportation}. The final step lies in performing a sequence of feed-forwards driven by homodyne measurement of the ancilla \cite{Gottesman}. Of those, the only nontrivial one is given by unitary $e^{i \lambda x^2}$, where the actual value of $\lambda$ depends on the measurement. This operation can be decomposed into a sequence of phase shift by $\phi_1$, squeezing with gain $g_f$ and another phase shift by $\phi_2$, where the parameters satisfy: $\tan \phi_1 \tan \phi_2 = -1$, $\tan \phi_1 = g_f$, and $(1-g_f^4)\cos\phi_1 \sin\phi_2 = 2 g_f \lambda$.
Adjusting the squeezing gain on the fly can be done by exploiting the universal squeezer \cite{MIO_theory, squeezing}, where the amount of squeezing is controlled by changing the ratio of the beam splitter, which can be done by a sequence of polarization beam splitter, polarization rotator, and another polarization beam splitter, where the rotator controls the splitting ratio. The nonlinear dependance of the feed-forward parameters on the measurement results requires a sufficiently fast data processing, but that too is available today \cite{adaptive}.

\textbf{Conclusion:} We have proposed an experimentally feasible way of deterministically achieving weak nonlinearity of the third order. The procedure effectively engineers the operation on a single photon level and then deterministically cut'n pastes the properties onto the target state. However, the nonlinearity is still composed only of a limited number of photon, so it can be faithfully applied only to target states which are sufficiently weak. Furthermore, since there is no such thing as free lunch, subsequent use of the transformed state in attempts to generate higher nonlinearities as per \cite{Lloyd99} requires higher and higher numbers of single photons used in the engineering.

The approach is not flawless. There are several sources of noise which can be simultaneously reduced only in the limit of infinitely small (read unobservable) interaction. This is due to the finite photon approximation of the cubic gate not being unitary and therefore not perfectly correctable by the unitary feed-forward. Nevertheless, we have shown that even with this noise a demonstration of decisively non-Gaussian high order quantum deterministic nonlinearity going well beyond classical attempts, based on higher order nonlinearity in the feed-forward loop, can be observed already now.

\textbf{Ackowledgements:}
This research has been supported by projects  MSM 6198959213, LC06007 and Czech-Japan Project ME10156 (MIQIP) of the Czech Ministry of Education. We also acknowledge grants P205/10/P319 of GA CR, EU grant FP7 212008 - COMPAS, and GIA, SCF, G-COE, FIRST, and APSA commissioned by the MEXT of Japan and SCOPE.

\end{document}